\documentclass[twocolumn,preprintnumbers,nofootinbib,prd]{revtex4}

\usepackage{graphicx}

\usepackage[hypertex]{hyperref}
\usepackage{amsmath, amssymb}
\newcommand{\vev}[1]{\langle {#1} \rangle}
\newcommand{\lsim}{\lesssim}

\newcommand{\co}{\mathcal{O}}
\newcommand{\gsim}{\gtrsim}
\newcommand{\beq}{\begin{equation}}
\newcommand{\eeq}{\end{equation}}

\newcommand{\lqcd}{\Lambda_{\tt QCD}}

\begin{document}

\pagestyle{plain}

\preprint{BNL-HET-07/6}

\title{Strong \boldmath $CP$, Up-Quark Mass, and the Randall-Sundrum Microscope}

\author{Hooman Davoudiasl\footnote{email: hooman@bnl.gov} }

\author{Amarjit Soni\footnote{email: soni@bnl.gov} }

\affiliation{Department of Physics, Brookhaven National Laboratory,
Upton, NY 11973, USA}


\begin{abstract}

In the Randall-Sundrum model, setting the ratio 
of up and down quark masses $m_u/m_d \ll 1$, 
relevant to the strong CP problem, does
not require chiral symmetry or fine-tuning, due to 
exponential bulk fermion profiles.  We point out that such geometric  
suppression of the mass of a fermion magnifies the 
masses of its corresponding Kaluza-Klein (KK) states. 
In this sense, these KK states act as ``microscopes" for probing
light quark and lepton masses.  In simple realizations, this hypothesis 
can be testable at future colliders, like the
LHC, by measuring the spectrum of level-1 KK fermions.
The microscope can then provide an experimental 
test for the vanishing of $m_u$ in the ultraviolet, 
independently of non-perturbative 
determinations, by 
lattice simulations or other means,
at hadronic scales.  
We also briefly comment on application of our microscope idea 
to other fermions, such as the electron and neutrinos.

\end{abstract}
\maketitle


The Standard Model (SM) has been extremely successful in describing observed
electroweak and strong phenomena.  In the SM, the Higgs 
condensate, $\vev{H} \simeq 250$~GeV,
sets the electroweak scale and QCD dynamics generates a scale $\Lambda_{\tt QCD}\sim
1$ ~GeV for strong ineractions.  The only other known interaction is 
gravity that
is assumed to be governed by the Planck scale $M_P \sim 10^{19}$~GeV.
Despite its remarkable success, the SM leaves a number of intriguing questions
unanswered.  Of these, three long-standing ones are the hierarchy problem, the
flavor puzzle, and the Strong $CP$ Problem (S$CP$P).  The Randall-Sundrum model
\cite{Randall:1999ee} was
initially proposed to resolve the first of the above problems.  Further work
showed that the second problem can also be addressed in this model 
\cite{Grossman:1999ra,Gherghetta:2000qt}.  

A vanishingly small up-quark mass $m_u$, 
an economical solution of the S$CP$P, can be 
naturally accomodated in the RS background 
\cite{otherS$CP$P}.  Our central observation is that, within 
the RS framework, 
{\it suppressing the mass of a zero mode (SM) fermion 
magnifies the masses of its KK states.}  We 
will refer to this phenomena as the ``RS microscope".     
Remarkably, the simplest realizations
of this RS resolution of the S$CP$P can be tested at multi-TeV-scale colliders, 
such as the LHC.  We note that this simple solution seems to be disfavored by 
the recent lattice results, 
taken at face value,
such as those in Ref.~\cite{Aubin:2004fs}.  
However, since establishing $m_u\approx 0$ at the weak-scale 
is of crucial importance, a direct 
weak scale collider test can be very useful.  This will 
avoid the subtle non-perturbative effects inherent in lattice 
and non-lattice results.  We will also outline how the microscope mechanism   
applies to other light fermions, such as the electron and neutrinos, 
and can give us insight into the RS flavor structure.

Before outlining the RS formalism relevant to our proposal, we would like to 
clarify some issues regarding the S$CP$P and $m_u$.  
To do this, we will first
give a brief description of the S$CP$P.  
In the SM, $CP$ violation in the strong sector is controlled by the parameter
\beq
{\bar \theta} = \theta + \arg \left(\det M_q\right)
\label{thetabar}
\eeq
that receives contributions from two seemingly unrelated sources:
$\theta$ which characterizes the gauge invariant QCD vacuum, and the quark mass
matrix $M_q$.  Therefore, absent any compelling physical principle,
the generic expectation is ${\bar \theta} \sim 1$.  However, this parameter
enters in the expression for the neutron electric dipole moment 
\cite{edmth} and experimental
bounds \cite{Baker:2006ts} require ${\bar\theta} \lsim 10^{-10}$.  
The reason for the smallness of
this parameter remains unknown, which is referred to as the S$CP$P.
Physics beyond the SM, such as the 
Peccei-Quinn mechanism \cite{Peccei:1977hh} 
may explain smallness of ${\bar \theta}$.
However, such proposals often require high scale
model-building whose key components cannot be directly
probed at colliders in the forseeable future.

Remarkably, there is a well-known 
solution to the S$CP$P problem that does not
require any physics beyond the SM, as suggested by
Eq.~(\ref{thetabar}): if there is a massless 
quark, ${\bar \theta}$ can be rotated
away \cite{MC}.  In fact, it suffices that 
the quark mass $m_q$ is such that
$\Im m_q \lsim 10^{-12}$~GeV \cite{Banks:1994yg}.  
Given the spectrum of the SM quarks, the only potential
choice is $q = u$, the up-quark.
Obviously, this could only be viewed as a resolution if
one can explain why $m_u$ is vanishingly small.  
The traditional way of addressing
this question is to assume that there is a 
chiral symmetry that prevents $u$ from
obtaining mass.  However, this symmetry will be 
anomalous under QCD interactions and
hence is not really a 
symmetry\footnote{This is also the case for the
Peccei-Quinn solution.  
}.  
One is then forced to think of this symmetry as accidental and
introduce the necessary structure at 
scales far above the SM that will result in the
desired outcome.  However, as we will explain below, the RS scenario 
accommodates a natural alternative 
mechanism for getting very small fermion masses.

At this point, we would like to address how setting the ratio of 
up and down quark masses $m_u/m_d\ll 1$, required 
to resolve S$CP$P, can be reconciled with the expectation 
that $m_u/m_d \lsim 0.3-0.5$ at
low energies of order $\lqcd$.   To see this, note that to solve the S$CP$P, 
it is sufficient to have $m_u \to 0$ at some scale above $\lqcd$.  
Below this scale, there is   
an additive {\it non-perturbative} contribution to 
the {\it real} part of $m_u$
\cite{Georgi:1981be,Choi:1988sy,Banks:1994yg}, due to QCD
instantons, that generates $m_u \neq 0$ at $\lqcd$, even if we set
$m_u = 0$ at the cutoff scale $\Lambda \sim 1$~TeV.  This
contribution has a form similar to that of the Kaplan-Manohar
ambiguity \cite{Kaplan:1986ru}, but is physically of a different
origin \cite{Banks:1994yg,Srednicki:2005wc}.  We have 
\cite{Georgi:1981be,Choi:1988sy,Banks:1994yg}
\beq
\frac{m_u}{m_d} \simeq \frac{1.15\, m_s}{12}\int_0^{\rho_0} 
d \rho\, \left[\frac{8 \pi^2}{g(\rho)^2}\right]^6 
e^{-\frac{8 \pi^2}{g(\rho)^2}},
\label{mu/md}
\eeq
where $m_s$ is the strange quark mass, $g(\rho)$ is the strong 
coupling constant at scale $1/\rho$, and $\rho$ is the instanton size.  
In the above formula, the integration has an infrared cutoff at $1/\rho_0$.  
For 3 flavors, the 1-loop expression
\beq
g^2/(8 \pi^2) = -1/[9 \ln(\rho \lqcd)]
\label{g2}
\eeq
is assumed in the above.  

Recent lattice results, like those of Ref.~\cite{Aubin:2004fs}, 
have constrained the efficiency of the above mechanism 
for non-perturbaive generation of $m_u$.  Ref.~\cite{Aubin:2004fs} gives, at 2~GeV,  
$m_u/m_d = 0.43 (0)(1)(8)$, where the errors are from statistics, simulation systematics, 
and electromagnetic effects, respectively. 
If we take $m_s = 100$~MeV, $\lqcd = 350$~MeV, and $1/\rho_0 = 2$~GeV, then we find
$m_u/m_d \simeq 2\times 10^{-3}$, which is far from the lattice result.  However, 
if we choose $m_s = 140$~MeV, $\lqcd = 500$~MeV, and $1/\rho_0 = 1.5$~GeV,  
we find $m_u/m_d \simeq 8\times 10^{-2}$, which is only a factor of 5 away from the 
lattice result.  

We do not advocate the latter values of parameters as more 
motivated; we only use them to show the sensitivity of the instanton result for 
$m_u/m_d$ to low energy parameters.  
Also note that the instanton density used in the above 
estimates can be subject to corrections by factors of 
${\cal} O(1)$, due to non-trivial 
QCD vacuum structure \cite{Choi:1988sy,Shifman:1979nz}.  Finally, we 
point out that the lattice results are reported at a scale which is well 
above the charm mass $m_c\simeq 1.3$~GeV.  However, the lattice 
simulations do not 
include charm, a light state at 2~GeV.  We 
do not speculate on the effect of including charm in the simulations.  Nonetheless, 
we expect that this omission is a source of systematic uncertainty and may 
change the current lattice estimates.  

Clearly Ref.~\cite{Aubin:2004fs} is a very serious and careful
lattice study of several of the important low energy 
constants and quark masses.  
We want to stress that our concern in here is only with regard 
to $m_u$. The notable point for $m_u$ 
is that $m_u/\lqcd \approx 5 \times 10^{-3}$. Thus even small systematic
effects may be quite relevant; charm quark above is just 
one example. In addition, use of NNLO-type  chiral perturbation
theory (rather than the full NNLO) fits, {\it e.g.},
may also be another
source of small sytematics that could be relevant to $m_u$.
We note also that in fact authors  of Ref.~\cite{Aubin:2004fs}
themselves were careful enough to caution that for
this delicate issue verification by another discretization
method is desirable. However, from our perspective it 
is just not clear that methods on~\cite{Aubin:2004fs},  
or off~\cite{Leutwyler:1989pn}, the lattice have all the rigor that
is needed to decide this critical issue.   

Furthermore, it also appears that any extraction of $m_u/m_d$ based on 
hadronic spectra is bound to be relevant at energies below 1~GeV.  If we set 
$1/\rho_0 = m_\rho \simeq 0.77$~GeV, in Eq.(\ref{mu/md}), using 
$m_s = 100$~MeV, $\lqcd = 350$~MeV, we find $m_u/m_d \simeq 0.34$, 
which is a reasonable value.  Given the sensitivity of theoretical 
results from the instanton calculations and possibility of 
small uncontrollable systematic effects, 
and given also that the possibilty of $m_u \approx 0$ is of
fundamental importance,
having a completely independent handle on $m_u/m_d$ at the weak scale, where 
it could resolve the S$CP$P, is very desirable.  
With all this in mind, in what follows, we will outline 
a possible collider test of this hypothesis, in the frameowrk of RS resolution 
of the gauge and flavor hierarchies.

The RS model explains
the ratio $\vev{H}/M_P \sim 10^{-17}$, 
by gravitationally red-shifting the 5-$d$ fundamental scale $M_5 \sim M_P$ down
to the TeV-scale, along the warped 5$^{\rm th}$ dimension.  This geomtery is based on
a slice of AdS$_5$, truncated by flat 4-$d$ boundaries often referred to as the Planck (UV)
and TeV (IR) branes.  The RS metric is given by \cite{Randall:1999ee}
\beq
ds^2 = e^{-2 \sigma} \eta_{\mu\nu} dx^\mu dx^\nu - r_c^2 d\phi^2,
\label{metric}
\eeq
where $\sigma = k r_c |\phi|$, $k$ is the 5-$d$ curvature scale, 
$r_c$ is the radius of compactification, $-\pi\leq\phi\leq\pi$,
and a $\mathbb Z_2$ orbifolding of the 5$^{\rm th}$ dimension is assumed.  

To solve the hierarchy problem, the Higgs is assumed to be localized near the TeV-brane,
where the reduced metric ``warps" $\vev{H}_5 \sim M_5$ down to the weak scale:
$\vev{H}_4 = e^{-kr_c \pi} \vev{H}_5$.  For $kr_c \approx 12$, we then get
$\vev{H}_{\tt SM} \equiv \vev{H}_4 \sim 1$~TeV.  Originally, it was proposed that all SM
content resides at the IR-brane.  However, as the cutoff scale in the 4-$d$
effective theory is also red-shifted to near the weak-scale, this would lead to
unsuppressed higher dimensional operators that result in large violations of experimental
bounds on various effects, such as those on flavor-changing neutral currents.  
This problem can be solved by realizing that points along the
warped 5$^{\rm th}$ dimension correspond to different effective 4-$d$ scales.
In particular, by localizing first and second generation fermions away from the IR-brane,
the effective scale that suppresses higher dimensional operators made up of these fields
is pushed to much higher scales \cite{Gherghetta:2000qt}.  
In the process of suppressing the dangerous operators,
this setup also leads to a natural mechanism for obtaining small fermion masses.

The above localization is achieved by introducing a 5-$d$ mass term in the bulk for
each fermion field \cite{Grossman:1999ra}.  
Let $c \equiv \mu/k$, where $\mu$ is the 5-$d$
mass of the fermion.  Each 5-$d$ fermion $\Psi$ has left- and right-handed
components $\Psi_{L,R}$ which can be expanded in Kaluza-Klein (KK) modes
\beq
\Psi_{L,R}(x, \phi) = \sum_{n=0}^{\infty}\psi_{L,R}^{(n)}(x)
\frac{e^{2 \sigma}}{\sqrt{r_c}} f^{(n)}_{L,R}(\phi).
\label{KKmodes}
\eeq
The KK wavefunctions $f^{(n)}_{L,R}$ are orthonormalized
\beq
\int d\phi \, e^\sigma f^{(m)}_{L,R}\, f^{(n)}_{L,R}
= \delta_{m n}.
\label{orthonormal}
\eeq
One can then show that the $n\neq 0$ modes are given by
$f^{(n)}_{L,R} \propto e^{\sigma/2} Z_{\frac{1}{2}\pm c} (z_n^{L,R})$
\cite{Grossman:1999ra}, where $Z_a = J_a + b_n Y_a$ 
is a linear combination of Bessel functions of order $a$, 
$z_n \equiv (m_n/k) e^\sigma$ and $m_n$ is the KK mass.  The zero-mode
wavefunction is given by
\beq
f^{(0)}_{L,R} = \frac{e^{\mp c \sigma}}{N^{L,R}_0},
\label{f0}
\eeq
with the normalization
\beq
N^{L,R}_0 = \left[\frac{e^{k r_c \pi(1\mp 2c)} - 1}{k r_c(1/2\mp c)}\right]^{1/2}.
\label{N0}
\eeq
   
In the SM, all $SU(2)$ doublets are left-handed, while the singlets are
right-handed.  Hence, one has to impose a ${\mathbb Z_2}$ parity on bulk
fermion fields so that only the doublets have left-handed zero modes and only
the singlets have right-handed zero modes.  
However, both the doublets and singlets have
left- and right-handed higher KK modes.  
Note that in our convention, the singlet (right-handed) and 
doublet (left-handed) zero mode wavefunctions are defined with 
opposite signs for mass paramaters $c^S$ and $c^D$, respectively.  This 
is consistent with our normalization conventions for the wavefunctions. 

The above profiles provide a mechanism for suppression of higher
dimensional operators involving light fermions, as mentioned before.
For example, consider a 4-fermion operator $O_{4\Psi}$
made up of $\Psi_i$, $i = 1,\ldots,4$, in the 5-$d$ theory
\beq
O_{4\Psi} = \frac{\Psi_1 \Psi_2 \Psi_3 \Psi_4}{M_5^3},
\label{O4Psi}
\eeq
with a coefficient of unity.  After dimensional reduction, 
the component of $O_{4\Psi}$
that is made up of the zero modes (SM fields) is given by 
\cite{Gherghetta:2000qt}
\beq
O_{4\psi} = \frac{\psi_1 \psi_2 \psi_3 \psi_4}{\Lambda_4^2},
\label{O4psi}
\eeq
where the 4-$d$ cutoff scale $\Lambda_4$ is given by
\beq
\Lambda_4 \equiv \sqrt{\frac{M_5^3\,k\,r_c^2
\left(4 - \sum_i a_i c_i\right)\prod_i N^\Psi_i}
{2\left[e^{(4 - \sum_i a_i c_i)kr_c\pi}-1\right]}}.
\label{Lam2}
\eeq
In deriving the above, the wavefunctions given by Eq.~(\ref{f0}) have been used,
and $a_i = \pm 1$ for $(L,R)$ choices, respectively.
The scale $\Lambda_4$ can be much higher than the weak scale
for light SM fermions.  

The above fermion profiles also lead to a natural scheme for SM fermion masses 
\cite{Grossman:1999ra,Gherghetta:2000qt}.
To see this, we will next examine the Yukawa interactions between fermions and the
Higgs field.  We will assume that the Higgs is on the IR-brane; this
is a very good approximation since the Higgs must be highly IR-localized.
Then, a typical Yukawa term in the 5-$d$ action will take the form
\beq
S^5_{\tt Y} = \int \!d^4x \; d\phi \sqrt{-g} \,\frac{\lambda_5}{k} H(x)
\Psi^D_L \Psi^S_R \delta(\phi-\pi),
\label{S5Y}
\eeq
where $\lambda_5 \sim 1$ is a dimensionless 5-$d$ Yukawa coupling and $\Psi^{D,S}$
are doublet left- and singlet right-handed 5-$d$ fermions, respectively.
After the rescaling $H \to e^{k r_c \pi} H$, the 4-$d$ action
resulting from Eq.(\ref{S5Y}) is
\beq
S^4_{\tt Y} = \int \!d^4x \; \sqrt{-g} \,\lambda_4 H(x)
\psi^{(D,0)}_L \psi^{(S,0)}_R + \ldots,
\label{S4Y}
\eeq
where the 4-$d$ Yukawa coupling for the corresponding
zero-mode SM fermion is given by \cite{Gherghetta:2000qt}
\beq
\lambda_4 = \frac{\lambda_5}{k r_c}\left[\frac{e^{(1-c^D+c^S)kr_c \pi}}
{N^{D,L}_0 N^{S,R}_0}\right],
\label{lam4}
\eeq
with $c^{D,S}$ denoting the 5-$d$ mass parameters for $\Psi^{D,S}$.  Thus,
in the quark sector, there are, in general, 9 different values for $c$'s: 3 for the
doublets and 6 for the singlets.  One can see that the exponential form of the
effective Yukawa coupling $\lambda_4$ can accommodate a large
hierarchy of values without the need for introducing unnaturally small
5-$d$ parameters.  

Thus, the RS background provides a geometric alternative
to the requirement of a chiral symmetry: 
fermion masses far below the weak scale can be naturally
obtained, without the need for tiny 5-$d$ Yukawa couplings.
In fact, this mechanism was originally proposed as a way of obtaining realistic
neutrino masses \cite{Grossman:1999ra}.  
If the $SU(2)$ singlet 5-$d$ up-quark, corresponding to
the right-handed up-quark of the SM, is localized near the UV-brane,
the resulting zero mode mass $m_u$ will be highly suppressed.
We will see that setting $m_u \lsim 10^{-12}$~GeV
at the weak scale in this context only amounts to choosing $c^S_u\sim 1$
and does not require fine-tuning of any underlying 5-$d$ parameters.
Therefore, it is natural to have
${\bar\theta} \lsim 10^{-10}$ in the RS model with bulk SM content!

A very interesting feature of the above scenario for obtaining $m_u \ll$~MeV is that
it is testable at multi-TeV-scale colliders, like the LHC.
The values of light quark masses $(u,d,s)$, 
as well as those of light leptons, are not observable in
high energy collider experiments.  However, in the simplest models based on
the above RS flavor-scheme, the spectrum of the KK states associated with SM
fermion of flavor $\alpha$ encodes the information 
about the bulk mass parameter $c_\alpha$, at
every KK level.  Hence, in principle, measuring all the first KK fermion
masses, and perhaps a few other quantities,
will provide a complete map of the bulk mass parameters $c_\alpha$.  This,
in turn can lead to a determination of the relative sizes of the ``hard'' masses
of zero mode quarks, at the weak scale and, in particular, the Lagrangian
parameter $m_u$. We will see in the following that a generic outcome of this
picture is a weak-singlet $u$-quark first KK state that is substantially heavier
than those associated with the other quarks.
\vskip 0.5cm
\begin{table}
[t]
\begin{tabular}{|l|c|c|c|c|}
\hline
${\rm Quarks}$
 & $c^D$
 & $c^S$
 & $m_q({\rm SM}) \; \; ({\rm GeV})\quad$
 & $m_q^{{\rm KK}}/m_g^{{\rm KK}}$ \\
\hline

$\left( \begin{array}{c}
u \\
d
\end{array} \right)$
 & 0.5
 &  $\left( \begin{array}{c}
-1.4 \\
-0.7
\end{array} \right)$
 &  $\left( \begin{array}{c}
3.5\times 10^{-14} \\
4.8\times 10^{-3}
\end{array} \right)$
 & 1.0, $\left( \begin{array}{c}
1.5 \\
1.1
\end{array} \right)$\\
\hline
$\left( \begin{array}{c}
c \\
s
\end{array} \right)$
 & 0.5
 &  $\left( \begin{array}{c}
-0.53 \\
-0.61
\end{array} \right)$
 &  $\left( \begin{array}{c}
1.2 \\
0.11
\end{array} \right)$
 & 1.0, $\left( \begin{array}{c}
1.0 \\
1.0
\end{array} \right)$\\
\hline
$\left( \begin{array}{c}
t \\
b
\end{array} \right)$
 & 0.4
 &  $\left( \begin{array}{c}
- \\
-0.52
\end{array} \right)$
 &  $\left( \begin{array}{c}
170.6 \\
4.1
\end{array} \right)$
 & 1.0, $\left( \begin{array}{c}
- \\
1.0
\end{array} \right)$\\
\hline
\end{tabular}
\caption{Sample values for a realistic set of SM bare 
quark masses.  The doublet and
singlet profile parameters are denoted by $c^D$ and $c^S$, respectively.
To get the top mass, a 5-$d$ Yukawa coupling $\lambda_5^t = 3.1$ has been assumed;
all other $\lambda_5 = 1$.  The resulting zero-mode
SM quark masses are given in GeV.  The last column is the ratio of the level-1
(Doublet, Singlet) KK quark masses to that of the KK gluon (gauge boson).
With $m_u$ set to a small value that resolves the S$CP$P, the level-1
singlet $u$-quark KK state is nearly $50\%$ heavier than any other of
its counterparts.  Mass splittings from 
KK-fermion Yukawa couplings have been ignored here.
}
\label{t1}
\end{table}

In passing, we point out that the S$CP$P can also be generated by 
higher dimensional operators in the RS
context.  To see this, consider the 5-$d$ operator
\beq {\cal O}_5 =
\frac{H H^\dag \varepsilon^{LMNPQ}G_{LM} G_{NP}\,\hat{n}_Q}{M_5^2}|_{\phi=\pi},
\label{O5}
\eeq
with $\hat{n}_Q = (0,0,0,0,1)$ and $G_{MN}$ the 5-$d$ gluon field strength.  
${\cal O}_5$ can be written at $\phi=\pi$, 
since 5-$d$ Lorentz invariance is broken by the IR-boundary.  In the 
4-$d$ effective theory, the above operator yields 
\beq {\cal O}_{4\tt CP} \sim 
\frac{H H^\dag G_{\mu\nu} \widetilde{G}^{\mu\nu}}{\Lambda^2},
\label{OCP4}
\eeq
where $\widetilde{G}^{\mu\nu}$ is the dual gluon field strength tensor,
leading to an effective $\theta$ parameter after electroweak
symmetry breaking, $\vev{H}\neq 0$: ${\cal O}_{4\tt CP} \to
\theta_{eff} G_{\mu\nu} \widetilde{G}^{\mu\nu}$.  
For $\Lambda \lsim 10$~TeV, 
we get $\theta_{eff} \gsim 6\times 10^{-4}$
which is at least 7 orders of magnitude too big.

The 5-$d$ SM scenario in the RS background
is subject to various experimental constraints,
including those from precision
electroweak and flavor data.  A number of models with
custodial $SU(2)_L\times SU(2)_R$ bulk symmetry have been
proposed to address these constraints 
\cite{Agashe:2003zs,Agashe:2004rs,Agashe:2006at}.  
In the following, we will
limit the scope of our study to SM fermion masses without specifying a
particular framework for such constraints.  We note that 
since we will treat $u_R$ and $d_R$ of the SM differently, 
models in which the weak singlets $(u, d)$ form an $SU(2)_R$ {\it doublet} 
cannot be used to solve S$CP$P, as proposed in our work.  
However, models with split doublets, such as that 
of Ref.~\cite{Agashe:2003zs} can be used in our setup.  We will not 
discuss the phenomenology of the extra exotics in these models, 
as they do not change the qualitative picture for the SM KK partners that
we present here.  This will suffice to
demonstrate our key observations.  
For a study of possible light exotic 
quarks in some warped scenarios 
see Ref.~\cite{Dennis:2007tv}.  For a detailed analysis of 
flavor physics in warped models with bulk custodial symmetry 
see Ref.~\cite{Agashe:2004cp}.  Here, we also note that
for $m_{KK} \sim 3$~TeV and for generic 
${\cal O}(1)$ phases in the RS bulk, the resulting neutron EDM is 
${\cal O}(20)$ too large \cite{Agashe:2004cp,Agashe:2004ay}.  
This feature may require 
additional model building or tuning at the 
${\cal O}(10^{-1})$ level.  However, we only 
focus on the ${\cal O}(10^{-10})$ tuning (S$CP$P) 
that is posed at the SM level, even if we decouple new physics.

We will adopt a simple setup in
which the Higgs field and the right-handed top quark are
localized on the IR-brane.  For simplicity and in 
order to have a transparent treatment of key features, 
we will assume a flavor-digonal bulk mass matrix.  
We will then choose the profile
parameters $c_\alpha$, $\alpha = 1, 2, \ldots, 8$, to get a
realistic SM fermion mass spectrum.  To fix the warp factor at
a specific value, we will assume that
$\vev{H}_5 = M_P \approx 1.2 \times 10^{19}$~GeV.
The relation $\vev{H}_4 = e^{-kr_c \pi} \vev{H}_5$
then yields $k r_c \simeq 12.2$.

In Table I, we provide a reference
set of values for $c_\alpha$ that result in realistic SM quark masses.
We have assumed a somewhat large value for the 5-$d$ top Yukawa coupling,
$\lambda_5^t = 3.1$.  All other 5-$d$ Yukawa couplings are set to unity, as
expected in a ``natural" model.  We only use the values of $c_\alpha$ as a
guide for our analysis.  In practice, there are
modest modulations in the profiles and
the values of $\lambda_5$, depending on the details of a warped scenario.
However, we do not expect large deviations from these typical numbers in
generic cases.  We have also ignored the effect of doublet-singlet 
KK-fermion Yukawa coulpings, which result in a small 
${\cal O}(\vev{H}/m_{KK})$ shift 
in the masses \cite{Agashe:2006iy}. We will comment on this later.  
We will not present a detailed numerical analysis, as
we are only interested in the most general aspects of this scenario for
addressing the S$CP$P.  
Next, we will outline the collider phenomenology of
SM KK fermions in our proposal.   

In the following, we will focus on 
KK quarks.  However, one expects that these
signals are accompanied by others, for example from the discovery of the
KK gluon \cite{KKgluon} or the 
KK graviton \cite{KKgraviton} of the 
RS model.  Thus, the value of $k r_c$ for the
warped background can in principle be assumed to be known.
Since the objects in which we are interested have color
charge, we will consider the quark-sector coupling to the gluon-sector
first.  In the rest of this work, by ``KK mode" we mean
the first level mode, unless otherwise specified.

In our setup, which we take as typical,
the zero modes couple as in the SM, by gauge invariance.  The same
is true for KK-quark coupling to the SM gluon.  However, the coupling of
quarks to KK gluons is modified compared to the SM.  The 5-$d$ action 
for the coupling of the bulk fermion $\Psi$ to the gauge field 
$A_M$ is given by
\beq
S_{\Psi A} = g_5\int \!d^4x \; d\phi V \left[V_l^M {\bar \Psi \gamma^l 
A_M \Psi}\right],
\label{PsiA}
\eeq
where $g_5$ is the 5-$d$ gauge coupling, 
$V$ is the determinant of the f\"{u}nfbein $V_l^M$, with 
$l=0,\ldots,3$, 
$V_\lambda^M = e^\sigma \delta^M_\lambda$, and $V_4^4 = -1$; 
$\gamma^l = (\gamma^\lambda, i \gamma^5)$.  The $A_\mu$ 
field can be expanded in KK modes as
\beq
A_\mu = \sum_{n=0}^\infty A_\mu^{(n)}(x) \frac{\chi^{(n)}(\phi)}{\sqrt{r_c}}.
\label{AKK}
\eeq
The gauge 
field KK wavefunctions are given by 
$\chi^{(n)}_A \propto e^{\sigma} Z_{1}(z_n)$ \cite{RSgauge}, 
subject to the orthonormality condition
\beq
\int_{0}^{\pi} d\phi \, \chi^{(n)} \chi^{(m)} 
= \delta_{m n}.
\label{ortho-gauge}
\eeq
Dimensional reduction of the action (\ref{PsiA}) 
then yields the couplings of the fermion and gauge KK towers.  
With our conventions, the 4-$d$ gauge coupling is given by 
$g_4 = g_5/\sqrt{2 \pi r_c}$.  

Because of UV-brane
localization, the coupling of two zero mode singlets to
a KK gluon is suppressed, by approximate 
volume suppression of order $(k r_c \pi)^{-1}$.
The singlet $b$-quark also has suppressed coupling to KK-gluons, due to
``accidental" orthogonality, since for $c \simeq 0.5$ the 
overlap integral is approximately of the form in Eq.~(\ref{ortho-gauge}), with 
one fermion replacing a constant gauge zero mode.
This feature also applies to the zero mode doublets (less so for doublet $b$), 
as they have $c \simeq 0.5$ as well.  The coupling to the singlet top-quark is
substantial, since it is an IR-brane field.  The couplings of light quark
singlet zero modes to their KK counter parts and a KK gluon is also suppressed,
since the KK-modes are IR-localized, but the singlets are UV-localized.  This
same coupling is ${\cal O}(1)$ for the singlet $b$-quark and 
possibly for the light quark doublets, as
they may not be highly UV-localized.

Let us now consider the coupling of the quarks to the Higgs.
Zero mode quarks couple to the Higgs with the usual Yukawa strength
which is mostly negligible, except for the top quark.
Next, we examine the couplings of a KK quark to a zero
mode quark and the Higgs.  The strength of this coupling
can affect the branching ratios of the KK quarks.  The coupling
of a singlet zero mode to a KK doublet and the Higgs is
suppressed to ${\cal O}(10^{-2})$ or less, 
due to the UV-localization of the zero mode.  
This coupling is, however, ${\cal O}(10^{-1})$ for the $b$-flavor.
The coupling of the singlet KK mode to a
doublet zero mode and the Higgs is also ${\cal O}(10^{-1})$, 
since none of these 
fields are UV-localized.  

Finally, we expect that coupling of 2 KK-quarks 
to the Higgs to be ${\cal O}(1)$, for the same
reason.  Numerically, using the values in Table I, we find this to be the
case.  Here, we note that this coupling generates an off-digonal element in the 
KK mass matrix, which leads to the mixing, as well as the splitting    
$\Delta_q \simeq \vev{H}/\sqrt{2}$,  
of the doublet and singlet 
KK quarks \cite{Agashe:2006iy}.  The lower bound from precision data on 
gauge KK mode masses is in the 2-3~TeV range \cite{Carena:2007ua}.  
Given that the quark KK modes are expected to be   
heavier than the gauge modes, the splitting 
$\Delta_q$ is a $(5-10)\%$ effect.  Also note that our key signal, 
a heavy (up-flavor) singlet KK mode, gives a $50\%$, or larger, effect.  
Hence, the off-diagonal 
couplings do not affect our conclusions significantly.  In the 
following discussion, we will then use the interaction basis, not the 
mass basis.  For 5-$d$ Yukawa couplings smaller than unity, the  
goodness of this approximation improves.

Given the above analysis, we generically expect that
the KK quarks are produced in pairs via
$q {\bar q}$ annihilation.  This production mechanism can provide
a multi-TeV reach for the KK quark modes, as deduced from collider studies
of gluino production in supersymmetric models 
\cite{Baer:1998sz,Baer:2003wx}.  However, depending on
the model, other production mechanisms may also be available.

In order to formulate a search strategy, we must consider the most
likely decay modes of each KK quark.  The doublet KK modes,
as well as the singlet $b$ KK mode, decay into a zero mode counterpart
and a KK gluon with relatively significant branching fractions.  This is
possible since all KK quarks are more massive than the KK gluon, and the
emission of zero mode quark is a negligible kinematic constraint.
However, such decay modes will be followed by the subsequent decay of the
KK gluon, which will then dominantly decay into right-handed top quarks,
in our setup.  Thus, for KK doublet quark $q^D_{KK}$,
we get the decay chain

\beq
q^D_{KK} \to g_{KK} (\to t_R {\bar t_R})\; q^D,
\label{qDKK}
\eeq
where $q^D$ is the zero mode doublet.  The above $t_R {\bar t_R}+ j$
signal can be used to reconstruct the mass of the original KK quark.  Note 
that the decay mode $q^D_{KK} \to H \, q^S$ is not relevant here, due to 
its suppressed rate. 

The decay mode in (\ref{qDKK}) will not be relevant 
for measuring the singlet KK quark masses due to its small 
branching fraction.  Since we would like to use the RS ``microscope" to
get a handle on $m_u$, we need to find a signal for the decay
of a singlet KK quark $q^S_{KK}$.  Here, the dominant mode is
\beq
q^S_{KK} \to q^D_{KK} H ,
\label{qSKK}
\eeq
where we 
generally expect the KK singlets to be heavier than the KK doublets and 
the above decay is allowed on-shell.  However, this mode will 
lead to a subsequent decay chain given in (\ref{qDKK}), which 
will require a complicated event reconstruction.  Hence, we may consider 
the subdominant, but more transparent decay mode
\beq
q^S_{KK} \to H \, q^D.
\label{qSKKII}
\eeq
Then, one can use the $H + j$ signal to measure the mass of the
parent KK singlet.  As noted before, we expect the corresponding mode to be 
suppressed for $q^D_{KK}$ and hence this is a signal for the singlet KK modes.  
Based on our scenario for getting $m_u \ll$~MeV,
we then expect that one of the reconstructed KK masses in this channel 
will be significantly larger than the rest, as can be seen from Table I.

To test the localization hypothesis, one must get a handle on the
relation between the KK masses and the zero mode masses.  For example,
by measuring the masses of the doublet and singlet KK $b$-quark, as
outlined above, one may infer the profile parameters $c^{D,S}_b$.  
Note that since the relevant decay modes for KK $b$-quarks 
in our setup are $b_{KK}^D \to t_R {\bar t}_R b^D$ and 
$b^S_{KK} \to H b^D$, $b$-tagging can single out the parent KK particle. 
As another reference point, we can take advantage of the microscopic 
effect in the lepton sector.  In particular, the electron is light enough
that the effect for its first KK mode may be significant.  Assuming 
$c^D \simeq 0.5$ for the doublet electron, we need $c^S \simeq -0.76$ to achieve 
a realistic electron mass.  In this case, the ratio of the 
first KK masses of the doublet and the singlet electrons to that of a 
KK gauge field will be 1.0 and 1.14, respectively.  

Such a mass difference 
is likely measurable.  This measurement may, 
however, require a multi-TeV $e^+ e^-$ collider.  
As in the quark sector, we expect that 
the $e^D_{KK}$ will decay into a zero mode 
doublet $e$ and a KK gauge field, say, the 
KK mode of a photon: $e^D_{KK} \to A_{KK}\, e^D$.  On the other hand, for 
the singlet KK mode, $e^S_{KK} \to H e^D$ is likely a reasonable 
decay mode to study.  Quantitaive statements about the above 
measurements in both the lepton and the quark sectors 
is outside the scope of this work.

Agreement of the above KK masses with the 
corresponding zero mode localization, 
as inferred from Eq.~(\ref{lam4}), will provide a test 
of the RS microscope and flavor generation mechanism.  
For the light quark flavors, in the 
simple small-doublet-singlet-mixing scenario we are considering, 
the events that reconstruct to $H + j$ 
will correspond to the singlets, while 
the doublets will decay into $t_R {\bar t}_R j$ and 
tightly cluster around a 
mass value corresponding to $c \simeq 0.5$.  
Once this is established, the observation
of a singlet KK quark (in the $H + j$ channel) with a substantially
larger mass than the rest, will point to extreme localization for that
flavor.  

This will provide collider evidence for a very small up quark mass, 
as it is the only flavor for which this assumption is feasible.
Since, in our scenario, the required mass 
difference between $u^S_{KK}$ and the other KK masses
is typically very large, the deduced $c_u$ parameters can yield the approximate
size of the $m_u$ in the effective Lagrangian.  If the deduced mass
is sufficiently small, it will provide evidence for a
resolution of the S$CP$P based on an``accidentally" 
tiny $m_u$.  This will be a probe of the high scale value of $m_u$, 
independently of the low energy lattice results.

In the above, we discussed the key collider phenomenology 
of the setup presented in our work.  However, we expect that 
the extreme localization of the singlet $u$-flavor will also have other 
possible ramifications.  For example, higher dimensional operators 
that contain the $u_R$ of the SM will be highly suppressed in this 
scenario, since the localization acts as an effective chiral 
symmetry.  We will not analyze here the consequences of the suppression
of these higher dimensional operators for low energy experiments.

Before closing, we would like to outline how the microscope mechanism 
can also be useful in other sectors; we will consider neutrino physics 
as an example.  Currently, neutrino oscillation experiments require 
at least two massive neutrinos, with 
$\delta m_\odot^2 \sim 10^{-4}$~eV$^2$ and $\delta m_{\rm atm}^2 
\sim 10^{-3}$~eV$^2$ \cite{Yao:2006px}, 
from solar and atmospheric data, respectively.  However, the third 
neutrino is still allowed to be massless, according to present data.  It 
would be very challenging to settle this question by future low energy 
measurements of the neutrino mass scale.  For example, 
the Katrin $\beta$-decay experiment will be sensitive down to 
$m_\nu \sim 0.2$~eV \cite{Valerius:2005aw}.  
We will next consider how the RS microscope can help.  

Let us assume that neutrinos 
are Dirac fermions.  In that case, a natural way to suppress neutrino 
masses would be by localizing singlet neutrino $\nu_R$ fields 
close to the UV-brane, as first 
suggested in Ref.~\cite{Grossman:1999ra}.  In order to address the data, 
we at least need to generate $m_\nu \sim \sqrt{\delta m_{\rm atm}^2} 
\sim 5 \times 10^{-2}$~eV.  This can be accommodated by choosing 
$c^D\simeq 0.5$ for the lepton doublet 
$(\nu_\ell, \ell)$, with $\ell = e, \mu, \tau$, and $c^S_{1,2} \sim -1.2$; 
the 5-$d$ Yuakawa interaction is
$(\lambda_5/k) H {\bar \nu_L} \nu_R$ with $\lambda_5=1$.  However, if the 
lightest state is substantially less massive, say, by a factor of 
$\co{(m_e/m_\tau)}$, we will need to consider $c^S_3 \sim -1.45$.  
In this case, the level-1 singlet KK modes corresponding to the heavy 
eigenstates are roughly $10\%$ lighter than the corresponding mode 
for the lightest neutrino.  This is a mass difference large enough that 
collider experiments can likely measure it.

We do not enter here into a detailed discussion 
of model-building or collider phenomenology 
of the neutrino sector.  However, we point out that KK 
modes of the singlet right-handed neutrinos can have 
substantial couplings to the SM, via their Yukawa interactions.  
Also, in models with bulk custodial symmetry, the right-handed 
KK neutrinos can be produced through their interactions with 
the $SU(2)_R$ gauge sector.  These interactions can make it possible 
for the $\nu_R$ KK states to be produced at the LHC or a multi-TeV 
$e^+e^-$ collider.  Thus, the microscope effect within the RS sceanrio 
can in principle shed 
light on the mass scale of neutrinos, which can be challenging 
for low energy experiments.  A more detailed study of the neutrino 
micrsocope is needed for more quantitative conclusions and lies outside 
the scope of this work.

In summary, the RS model provides a natural setting for 
vanishingly small $m_u$, by fermion localization in the bulk.  This 
can be a potential solution to the S$CP$P.  An interesting 
feature of this scenario is that it is testable at a multi-TeV-scale collider.  
This is because the KK quark corresponding to the 
the right-handed up-quark in the SM will get a substantially larger 
mass than the other KK quarks, at the same level.  In principle, one can 
measure the spectrum of the level-1 KK quarks, say, at the LHC and 
deduce that one of the light quarks has a very tiny mass.  
This would be a probe of $m_u/m_d$ that is  
complementary and independent of the ongoing non-perturbative 
lattice simulations.  The enhancement of KK state masses can also shed light 
on the spectrum of other light fermions, 
such as a Dirac neutrino which is much lighter than the other two.  
Here again, collider data can provide a new handle on the neutrino mass 
scale, independently of low energy $\beta$-decay experiments.

\acknowledgments

We would like to thank M. Creutz, J. Hewett, 
T. Krupovnickas, W. Marciano, G. Perez, 
and F. Petriello for discussions.  
This work was supported in part by the United States Department of
Energy under Grant Contracts DE-AC02-98CH10886.



\begin{thebibliography}{99}

\bibitem{Randall:1999ee}
  L.~Randall and R.~Sundrum,
  Phys.\ Rev.\ Lett.\  {\bf 83}, 3370 (1999)
  [arXiv:hep-ph/9905221].

\bibitem{Grossman:1999ra}
  Y.~Grossman and M.~Neubert,
  Phys.\ Lett.\  B {\bf 474}, 361 (2000)
  [arXiv:hep-ph/9912408].


\bibitem{Gherghetta:2000qt}
  T.~Gherghetta and A.~Pomarol,
  Nucl.\ Phys.\  B {\bf 586} (2000) 141
  [arXiv:hep-ph/0003129].

\bibitem{otherS$CP$P}
There are a variety of other models that discuss 
S$CP$P within the RS model.  However these models do not 
consider the vanishing of $m_u$ and 
rely on extra structure to address the S$CP$P.  
See, for example Ref.\cite{warpedS$CP$P}.

\bibitem{warpedS$CP$P}
  H.~Collins and R.~Holman,
  Phys.\ Rev.\  D {\bf 67}, 105004 (2003)
  [arXiv:hep-ph/0210110].
  A.~Fukunaga and K.~I.~Izawa,
  Phys.\ Lett.\  B {\bf 562}, 251 (2003)
  [arXiv:hep-ph/0301273];
  K.~w.~Choi,
  Phys.\ Rev.\ Lett.\  {\bf 92}, 101602 (2004)
  [arXiv:hep-ph/0308024];
  T.~Flacke, B.~Gripaios, J.~March-Russell and D.~Maybury,
  JHEP {\bf 0701}, 061 (2007)
  [arXiv:hep-ph/0611278].

\bibitem{Aubin:2004fs}
  C.~Aubin {\it et al.}  [MILC Collaboration],
  Phys.\ Rev.\  D {\bf 70}, 114501 (2004)
  [arXiv:hep-lat/0407028].


\bibitem{edmth}
  V.~Baluni,
  Phys.\ Rev.\  D {\bf 19}, 2227 (1979);
R.~J.~Crewther, P.~Di Vecchia, G.~Veneziano and E.~Witten,
Phys.\ Lett.\  B {\bf 88}, 123 (1979)
[Erratum-ibid.\  B {\bf 91}, 487 (1980)]
K.~Kawarabayashi and N.~Ohta,
  Prog.\ Theor.\ Phys.\  {\bf 66}, 1789 (1981); 
  M.~Pospelov and A.~Ritz,
  Phys.\ Rev.\ Lett.\  {\bf 83}, 2526 (1999)
  [arXiv:hep-ph/9904483].

\bibitem{Baker:2006ts}
  C.~A.~Baker {\it et al.},
  Phys.\ Rev.\ Lett.\  {\bf 97}, 131801 (2006)
  [arXiv:hep-ex/0602020].


\bibitem{Peccei:1977hh}
  R.~D.~Peccei and H.~R.~Quinn,
  Phys.\ Rev.\ Lett.\  {\bf 38}, 1440 (1977);
  Phys.\ Rev.\ D {\bf 16}, 1791 (1977).

\bibitem{MC}

For comments on this point of view, see 
Refs.~\cite{Creutz:2004fi,Srednicki:2005wc}.

\bibitem{Creutz:2004fi}
  M.~Creutz,
  Phys.\ Rev.\ Lett.\  {\bf 92} (2004) 162003.

\bibitem{Srednicki:2005wc}
  M.~Srednicki,
  Phys.\ Rev.\ Lett.\  {\bf 95}, 059101 (2005)
  [arXiv:hep-ph/0503051].


\bibitem{Georgi:1981be}
  H.~Georgi and I.~N.~McArthur, HUTP-81/A011 (1981).

\bibitem{Choi:1988sy}
  K.~Choi, C.~W.~Kim and W.~K.~Sze,
  Phys.\ Rev.\ Lett.\  {\bf 61}, 794 (1988).

\bibitem{Banks:1994yg}
  T.~Banks, Y.~Nir and N.~Seiberg,
  arXiv:hep-ph/9403203.


\bibitem{Kaplan:1986ru}
  D.~B.~Kaplan and A.~V.~Manohar,
  Phys.\ Rev.\ Lett.\  {\bf 56}, 2004 (1986).



\bibitem{Shifman:1979nz}
  M.~A.~Shifman, A.~I.~Vainshtein and V.~I.~Zakharov,
  Nucl.\ Phys.\  B {\bf 165}, 45 (1980).

\bibitem{Leutwyler:1989pn}
  H.~Leutwyler,
      Nucl.\ Phys.\  B {\bf 337}, 108 (1990).



\bibitem{Agashe:2003zs}
  K.~Agashe, A.~Delgado, M.~J.~May and R.~Sundrum,
  JHEP {\bf 0308}, 050 (2003)
  [arXiv:hep-ph/0308036].

\bibitem{Agashe:2004rs}
  K.~Agashe, R.~Contino and A.~Pomarol,
  Nucl.\ Phys.\  B {\bf 719}, 165 (2005)
  [arXiv:hep-ph/0412089].

\bibitem{Agashe:2006at}
  K.~Agashe, R.~Contino, L.~Da Rold and A.~Pomarol,
  Phys.\ Lett.\  B {\bf 641}, 62 (2006)
  [arXiv:hep-ph/0605341].

\bibitem{Dennis:2007tv}
  C.~Dennis, M.~K.~Unel, G.~Servant and J.~Tseng,
  arXiv:hep-ph/0701158.

\bibitem{Agashe:2004cp}
  K.~Agashe, G.~Perez and A.~Soni,
  Phys.\ Rev.\  D {\bf 71}, 016002 (2005)
  [arXiv:hep-ph/0408134].

\bibitem{Agashe:2004ay}
  K.~Agashe, G.~Perez and A.~Soni,
  Phys.\ Rev.\ Lett.\  {\bf 93}, 201804 (2004)
  [arXiv:hep-ph/0406101].

\bibitem{Agashe:2006iy}
  K.~Agashe, A.~E.~Blechman and F.~Petriello,
  Phys.\ Rev.\  D {\bf 74}, 053011 (2006)
  [arXiv:hep-ph/0606021].


\bibitem{KKgluon}
  K.~Agashe, A.~Belyaev, T.~Krupovnickas, G.~Perez and J.~Virzi,
  arXiv:hep-ph/0612015; 
  B.~Lillie, L.~Randall and L.~T.~Wang,
  arXiv:hep-ph/0701166.


\bibitem{KKgraviton}
  H.~Davoudiasl, J.~L.~Hewett and T.~G.~Rizzo,
  Phys.\ Rev.\ Lett.\  {\bf 84}, 2080 (2000)
  [arXiv:hep-ph/9909255];
  H.~Davoudiasl, J.~L.~Hewett and T.~G.~Rizzo,
  Phys.\ Rev.\  D {\bf 63}, 075004 (2001)
  [arXiv:hep-ph/0006041]; 
  A.~L.~Fitzpatrick, J.~Kaplan, L.~Randall and L.~T.~Wang,
  arXiv:hep-ph/0701150;
  K.~Agashe, H.~Davoudiasl, G.~Perez and A.~Soni,
  arXiv:hep-ph/0701186.



\bibitem{RSgauge}
  H.~Davoudiasl, J.~L.~Hewett and T.~G.~Rizzo,
  Phys.\ Lett.\  B {\bf 473}, 43 (2000)
  [arXiv:hep-ph/9911262]; 
  A.~Pomarol,
  Phys.\ Lett.\  B {\bf 486}, 153 (2000)
  [arXiv:hep-ph/9911294].


\bibitem{Carena:2007ua}
  M.~Carena, E.~Ponton, J.~Santiago and C.~E.~M.~Wagner,
  arXiv:hep-ph/0701055.

\bibitem{Baer:1998sz}
  H.~Baer, C.~h.~Chen, M.~Drees, F.~Paige and X.~Tata,
  Phys.\ Rev.\  D {\bf 59}, 055014 (1999)
  [arXiv:hep-ph/9809223].

\bibitem{Baer:2003wx}
  H.~Baer, C.~Balazs, A.~Belyaev, T.~Krupovnickas and X.~Tata,
  JHEP {\bf 0306}, 054 (2003)
  [arXiv:hep-ph/0304303].

\bibitem{Yao:2006px}
  W.~M.~Yao {\it et al.}  [Particle Data Group],
  J.\ Phys.\ G {\bf 33} (2006) 1.

\bibitem{Valerius:2005aw}
  K.~Valerius  [KATRIN Collaboration],
  PoS {\bf HEP2005}, 166 (2006).




\end{thebibliography}
\end{document}